\documentclass[%
 aps,
 prl,%
 amsmath,amssymb,
 reprint,
showkeys,
]{revtex4-2}

\usepackage[T1]{fontenc}
\usepackage[utf8]{inputenc}
\usepackage{amsmath}
\usepackage{amssymb}
\usepackage{graphicx}
\usepackage{dcolumn}
\usepackage{refstyle}
\usepackage{multirow}
\usepackage[
  citecolor=blue,
  colorlinks,
  linkcolor=blue,
  urlcolor=blue,
]{hyperref}
\usepackage[dvipsnames]{xcolor}
\usepackage{dcolumn}
\usepackage{bm}

\begin{document}

\preprint{APS/123-QED}

\title{Effect of externally deposited nanoscale heterogeneities in thin polymer films on their adhesion behavior}

\author{Chiranjit Majhi}
\affiliation{
	Department of Physics, Indian Institute of Technology Kanpur, Kanpur - 208016,	India
}
\author{Bidisha Bhatt}
\affiliation{
	Department of Physics, Indian Institute of Technology Kanpur, Kanpur - 208016,	India
}
\author{Shivam Gupta}
\affiliation{
	Department of Physics, Indian Institute of Technology Kanpur, Kanpur - 208016,	India
}
\author{Krishnacharya Khare}
\email{kcharya@iitk.ac.in}
\affiliation{
	Department of Physics, Indian Institute of Technology Kanpur, Kanpur - 208016,	India
}

\begin{abstract}
Adhesion between two surfaces depend on the chemical and the mechanical properties of both the materials. However, heterogeneities (surface or bulk) affect the adhesion between the two surfaces tremendously. In this work, we study the role of externally deposited nanoscale heterogeneities on their adhesion behavior. Silica nanoparticles in the bulk polydimethylsiloxane (PDMS) polymer matrix act as external heterogeneities, which subsequently affect their adhesion. The nanoscale heterogeneities change the polymer environment locally, which subsequently modify its mechanical properties, e.g. elastic modulus, as a result, it affects its adhesion behavior. We observe that the adhesion behavior vary in nonlinear manner with increasing nanoparticle concentration. Therefore, probing these heterogeneities may allow us to understand the emergence of processing-induced deviations and the role of external defects in the macroscopic properties of a polymer.
\end{abstract}

\maketitle

\section{Introduction}
Polymers are made with many repeating monomer units in a predefined manner. Depending on the number of monomers, a polymer has a fixed molecular weight and viscosity~\cite{fox1948viscosity}. Small molecular weight polymers behave in a manner similar to Newtonian fluids as the molecular chains stay linear in equilibrium as well as under external shear~\cite{moore2000molecular}. However, large molecular weight polymers behave as non-Newtonian fluids since the molecular chains are usually entangled. Such polymers show highly non-linear mechanical and flow properties~\cite{dukhin2020rheology}. 

The heterogeneities can also be of two types; internal, due to residual stress and/or polymer chain conformation, and external, due to impurities. The entangled sites behave as local nanoscale heterogeneities, which subsequently affect the properties of the polymer~\cite{chandran2017time, thomas2011nonequilibrium, chandran2019processing}. 

It is well known that many macroscopic properties of polymers, such as glass transition temperature, elastic modulus, rheological behavior, and rupture probability, depend on the microscopic heterogeneities in the polymer~\cite{singh2020glass, chandran2019processing, liu2022dynamic, mahmud2020enhancing, alesadi2020understanding, chung2009quantifying}. However, processing-induced nanoscale heterogeneities in polymer films remain largely unexplored, mainly because of the experimental difficulty in accessing the appropriate length scales. 

The molecular deformations which appear from the nonequilibrium processing pathway may result in spatially inhomogeneous systems at the length scales of a few coils, i.e., of the order of a few nm to few tens of nm. Experiments indicate that such nonequilibrium conformations and, thereby, resulting heterogeneities give rise to structural, dynamical, and mechanical properties that differ strongly from those in thermodynamic equilibrium\cite{chandran2019processing, rubinstein2003polymer, de1979scaling, reiter2005residual}. These resulting heterogeneities also reflect via interfacial properties like interfacial adhesion ~\cite{chazot2019understanding}. Highly selective adhesion can be achieved between surfaces by patterning them with ripples or patterns~\cite{vajpayee2011adhesion}. It becomes more complex when uniform dispersion of nanoparticles is added to the polymeric material. 

One way to investigate the role of nanoscale heterogeneities is via its manifestation in bulk mechanical behavior~\cite{qiao2019structural}.
Nanocomposites of polydimethylsiloxane (Dow Corning, USA)-$\mathrm{SiO_2}$ hydrophilic nanoparticles (20 nm) (NP) were used as a model system in the study.

\section{Materials and Methods}
\begin{figure*}[ht!]
\centering
\includegraphics[width=0.8\textwidth]{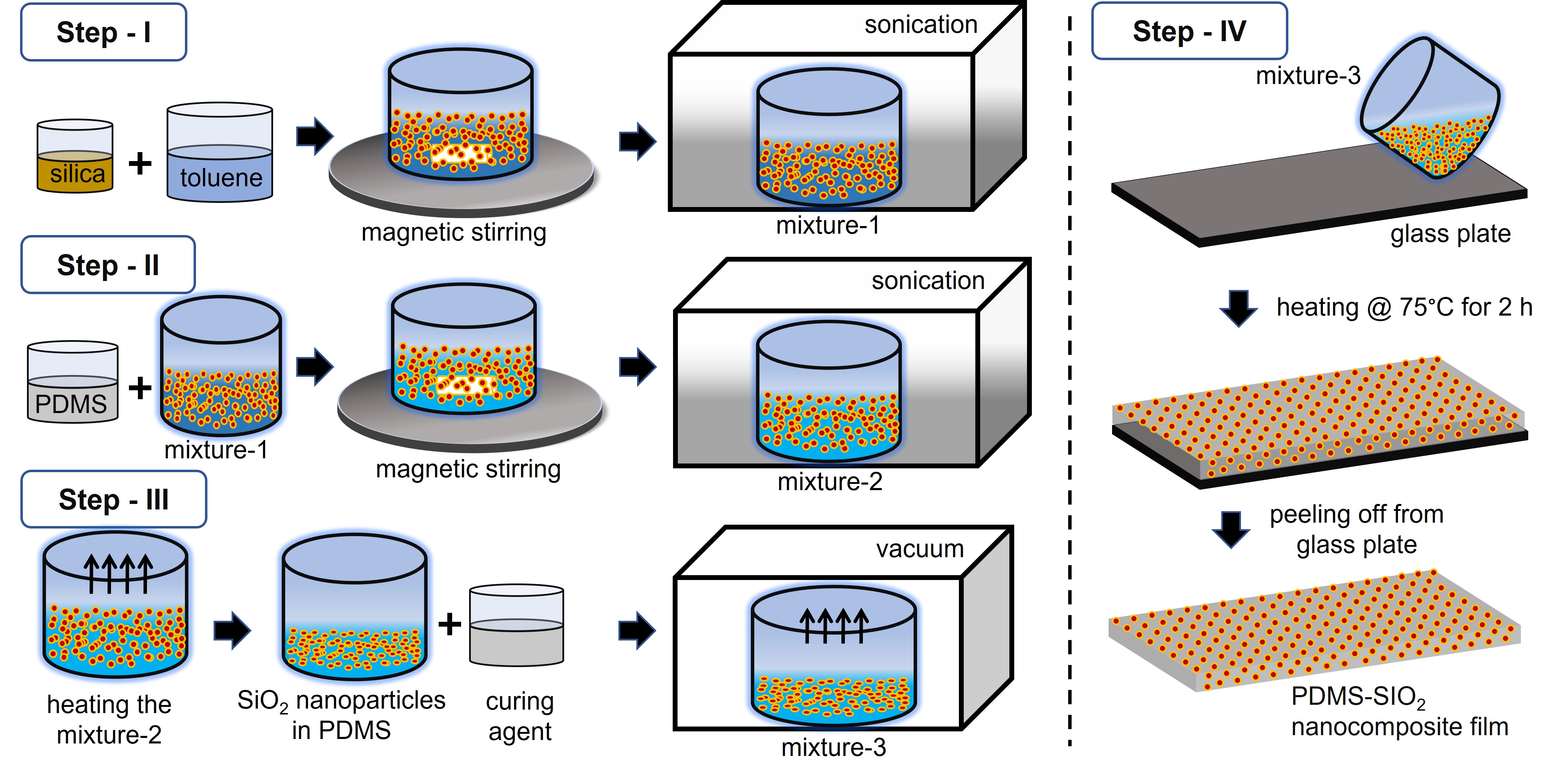}
\caption {Procedures for making PDMS-$\mathrm{SiO_2}$ nanocomposites. (step-I) silica nanoparticles dispersed in toluene using magnetic stirring and sonication; (step-II) silica nanoparticles mixed with PDMS base using magnetic stirring and sonication; (step-III) evaporation of toluene and mixing of curing agent with PDMS-$\mathrm{SiO_2}$ nanocomposite. The bubbles were removed in the desiccator at low pressure; (step-IV) pouring of PDMS-$\mathrm{SiO_2}$ nanocomposites system on a glass plate and heating for curing. PDMS-$\mathrm{SiO_2}$ nanocomposite sheets were collected by peeling off from the glass plate. }
\label{fig:1}
\end{figure*}
The polydimethylsiloxane (PDMS) and the PDMS-$\mathrm{SiO_2}$ nanocomposite sheets were prepared separately. To prepare the PDMS sheets, a standard procedure was followed. First, the PDMS base was mixed with the curing agent (Dow Corning, USA) at a ratio of 10:1, respectively. During the process of stirring, air bubbles were formed. The bubbles were removed in the desiccator at an ambient pressure of $30$ mmHg for $1$ h. The mixture was then poured between two glass plates and cured at $75 ^{\circ}$ C for 2 h. Compared to the procedure for the preparation of PDMS sheets, the procedure for the nanocomposites was a little laborious. PDMS-$\mathrm{SiO_2}$ nanocomposites were prepared with the use of multiple steps, as shown in Fig. \ref{fig:1}. Firstly, a solution was prepared by mixing the nanoparticles in toluene (Loba Chemie, India) using a magnetic stirrer (250 rpm) for 4 h, followed by ultrasonication for 4 h.  Next, the solution was mixed with the PMDS base (Sylgard 184), with the concentration of nanoparticles varying from 0.1-20 (w/w) in the PDMS base. The prepared mixture was stirred and ultrasonicated for 6 h. Subsequently, the volatile solvent was evaporated at $70^{\circ}$ C temperature for a few hours to evaporate the toluene in a chemical hood thoroughly. During evaporation, the mixture was constantly stirred to prevent the agglomeration of the nanoparticles. After the evaporation of toluene, the curing agent was added to the mixture with a 10:1 (w/w) ratio of PDMS and the curing agent, respectively. Again, the process of stirring led to the formation of air bubbles. The bubbles were removed using a similar procedure as reported before. The nanocomposite sheets were prepared by pouring the mixture between two glass plates and curing at $75 ^{\circ}$C for 2 h. After the curing of PDMS and PDMS-nanocomposite sheets (thickness $\sim 1$ mm), they were cut into 20 mm  $\times$ 35 mm area and were subsequently peeled off from the glass plate for further analysis. The prepared sheets were analyzed optically and subsequently used for adhesion measurements. 

\section{Results and Discussion}
The quality of the prepared samples was observed using an upright microscope (BX-51, Olympus, Japan). To confirm the homogeneous dispersion of the nanoparticles, we observed the transmittance of the samples in white light using a stereo-zoom microscope (RSM-9, Radical, India) in transmission mode. The transmittance was analyzed using open source software, imagej~\cite{schneider2012nih, schindelin2015imagej, collins2007imagej, gomez2021deepimagej}. 
\begin{figure*}[ht!]
\centering
\includegraphics[width=0.7\textwidth]{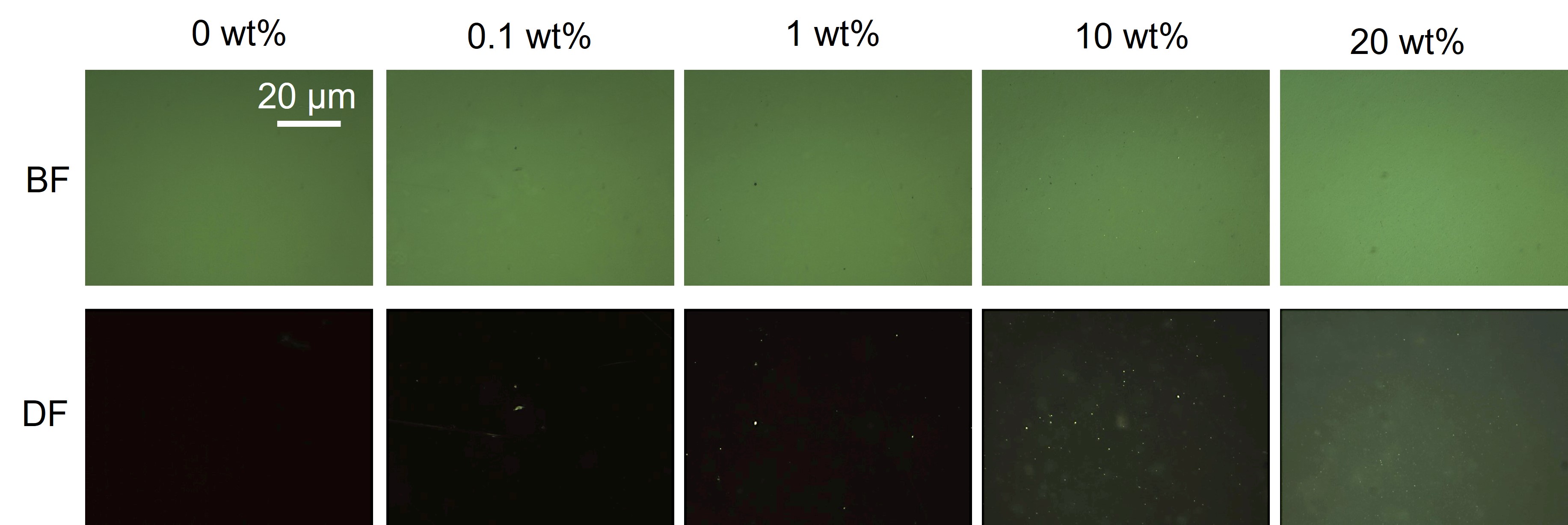}
\caption {Optical images of PDMS-$\mathrm{SiO_2}$ nanocomposites for varying concentration of nanoparticle (0-20\%) (w/w) in bright field (BF) and dark field (DF) microscopy in reflection mode. Different modes have been captured to demonstrate homogeneous nanoparticle distribution at the surface of the samples. }
\label{fig:2}
\end{figure*}

\begin{figure}[h!]
\centering
\includegraphics[width=0.5\textwidth]{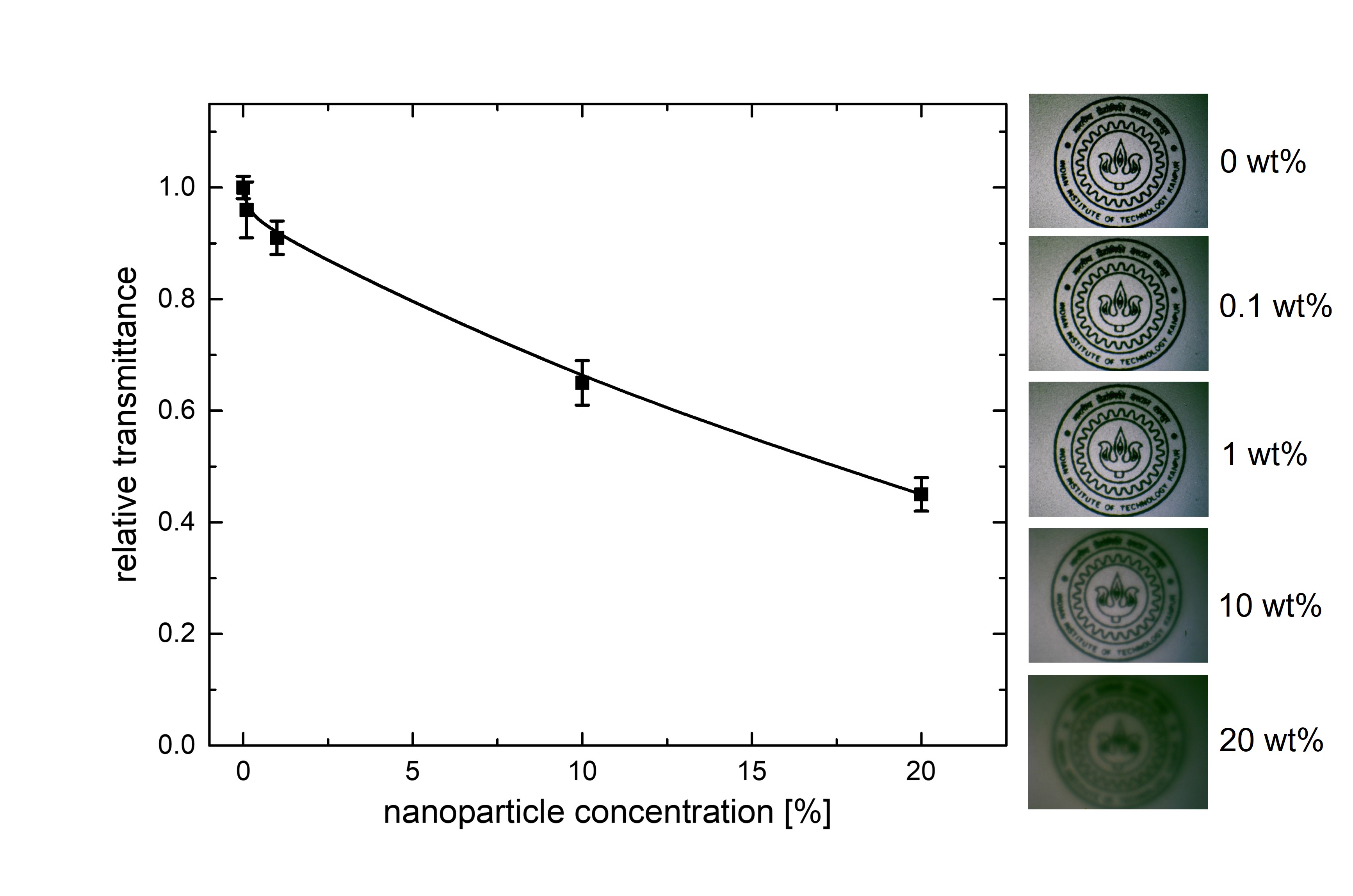}
\caption {Transmittance of PDMS nanocomposite films as a function of silica nanoparticle concentration.}
\label{fig:3}
\end{figure}

First, the quality of the prepared sheets with varying nanoparticle concentrations in PDMS was ensured. The step is crucial since the heterogeneous dispersion of the nanoparticles can lead to misinterpretation of the observed results. The samples were imaged in both bright field (BF) and dark field (DF) modes in an upright microscope, as shown in  Fig. \ref{fig:2}. For reference, we used the PDMS sheet without nanoparticles and referred to it as 0 wt\%. From the images in BF and DF, it is clear that the nanoparticles do not heterogeneously disperse. Further information about the dispersion and the distribution can be understood from the transmittance results shown in Fig. \ref{fig:3}. From the plot of the transmittance of the light as a function of the concentration of the nanoparticles, it is observed that the light intensity decreases as the concentration is increased. The behavior is also complemented by the optical images shown in the inset of Fig. \ref{fig:3}. The reason for such behavior is the dispersion of light by the dispersed nanoparticles, which increases as more nanoparticles are added to the PDMS network. 

\begin{figure*}[ht!]
\centering
\includegraphics[width=1\textwidth]{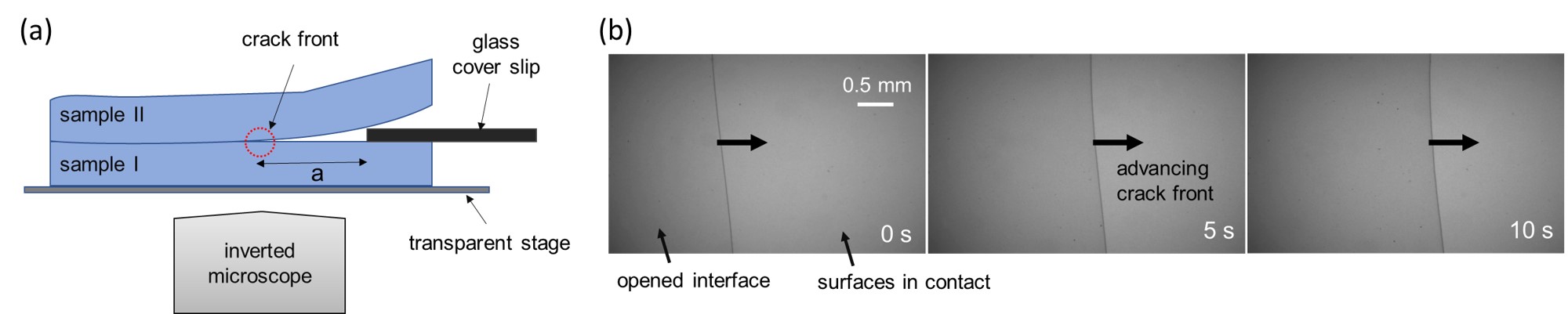}
\caption {(a) Experimental set-up for measuring adhesion. (b) Optical microscopic images of two flat surfaces show a crack advancing through the interface. The black arrow points to the face of the open crack showing some surface features; the block arrows point to the region of complete contact.}
\label{fig:4}
\end{figure*}

To study the crack length, we adopted the following methodology. First, a thin glass coverslip ($\sim 0.15 \mathrm{mm}$ thick) was placed between the two surfaces at the endpoints as shown in Fig. \ref{fig:4}. 
Subsequently, the two sheets were pressed against each other while keeping the glass coverslip sandwiched between the surfaces. To achieve good contact between the two samples, it is essential to apply sufficient pressure. In our experiment, we used a smooth glass rod of diameter 5 mm for pressing the two sheets to achieve good contact. Next, the applied pressure was released, and the crack started proceeding due to the spontaneous opening of the interface.
The crack propagation was analyzed using an inverted microscope (IX 73, Olympus, Japan). With the help of Tracker, an open source software ~\cite{brown2009innovative, moraru2021distance, rodrigues2013teaching, wee2012using}, we analyzed the propagation of the crack tip as a function of time and the   equilibrium crack length (the point at which crack length gets arrested) as a function of the concentration of nanoparticles. 

\begin{figure}[h!]
\centering
\includegraphics[width=0.45\textwidth]{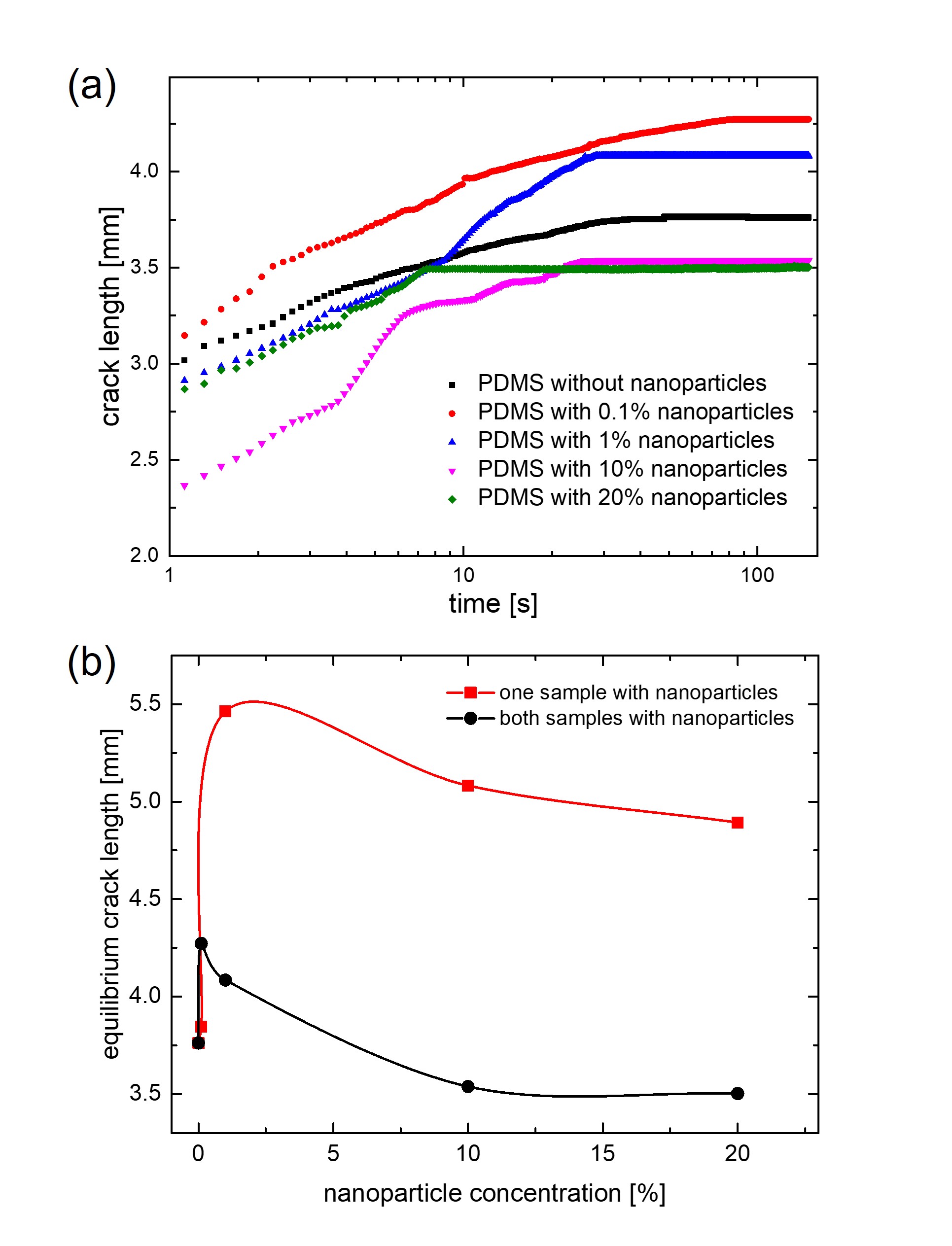}
\caption {(a) Crack length as a function of time after insertion of a wedge into the interface between two surfaces. Each data set is for a set of identical surfaces with varying nanoparticles concentration (0-20\%) w/w. From the plateau region of each data set, we can measure the equilibrium crack length. (b) Equilibrium crack length as a function of nanoparticles concentration (0-20\% w/w). One data set is for a set of identical surfaces, and another is for non-identical surfaces(one pure PDMS and one PDMS with filler particles).}
\label{fig:6}
\end{figure}

To understand the effect of nanoparticle concentration on crack tip propagation, we rely on the theory for mode $I$ crack propagation ~\cite{lawn1993fracture}. Since, the roughness of the surface is very small compared to any other dimension, such as the crack length or sample size, we focus our attention on a semi-infinite crack tip subjected to remote mode $I$ loading. In the absence of any weak undulating interface in the elastic medium, when normal tensile stress is applied, the crack expands perpendicularly to the loading, as the energy release rate is maximum in that direction. This is the same direction in which a crack would expand if the strip had a flat interface. In the event of weak contact with undulations, the direction of crack propagation may vary as the crack grows. Therefore, the energy release rate for an undulating interface will be lower than that of a straight crack, at least at some point along its trajectory. This can be demonstrated by calculating the energy release rate for the crack when the direction of the maximum energy release rate makes an angle $\theta$ with the direction of crack growth. Assuming that the open crack face behind the crack tip is flat, the stress intensity factors at the crack tip are~\cite{lawn1993fracture},
\begin{align}
      K_{I}^{'} = K_I f_{\theta \theta}^I \\
      K_{II}^{'} = K_I f_{r \theta}^I
      \label{eq1}
\end{align}
where $K_I$ is the stress intensity factor for mode $I$ and applied to the crack opening mode $I$. The functions $f_{\theta \theta}^I$ and $f_{ r \theta}^I$
are given by 
\begin{align}
    \begin{split}
        f_{\theta \theta}^I = {\cos}^3 {(\theta/2)} \\
        f_{r \theta}^I = {\sin}{(\theta/2)}\:  {\cos}^2 {(\theta/2)}
    \end{split}
    \label{eq3}
\end{align}
On combining the above equations, the energy release rate $\mathrm{G_r}$ defined as $G_r = (K_{I}^{'2} + K_{II}^{'2})/{E^{*}}$ is given by
\begin{equation}
    G_r = \frac{K_{I}^{2}}{E^{*}}\: {\cos}^4{\left(\frac{\theta}{2}\right)}
    \label{eq4}
\end{equation}
where $\mathrm{E^{*} = E / (1-\nu^2)}$ is the plane elastic modulus ($E$ is Young's modulus and $\mathrm{\nu}$ is Poisson's ratio of the substrate). From equation (\ref{eq4}), we can see that the energy release rate reduces by a factor ${\cos}^4 (\theta/2)$ and the energy release rate along a strip will be minimum at a maximum value of $\theta$ gives as 
\begin{equation}
    G_{min} = W_{\textnormal{ad}} \: {\cos}^4 \left(\frac{\theta_{max}}{2}\right)
    \label{eq5}
\end{equation}
where ${W_{\textnormal{ad}}}$ is the work of adhesion of the interface. 
Waters and Guduru observed an increase in the effective work of adhesion at an interface under mixed-mode loading~\cite{waters2010mode}. So, we can express the energy release rate as a function of the phase angle of loading, $\Psi$, and crack velocity, $v$ given as 
\begin{align}
    G_r &= W_{\textnormal{ad}}^r (\Psi, v)
\end{align}    
where $W_{\textnormal{ad}}^r$ is the interfacial work of adhesion and $ \tan (\Psi) = \frac{K_{II}^{'}}{K_{I}^{'}} = \tan \left( \frac{\theta}{2}\right) \Rightarrow \Psi =  \frac{\theta}{2} $
Replacing the phase angle in terms of $\theta$ we get
\begin{equation}
    G_r = W_{\textnormal{ad}}^r ( \frac{\theta}{2}, v)
    \label{eq6}
\end{equation}
Equation (\ref{eq6}) predicts that for a flat contact the energy release rate in quasistatic crack propagation is related to the work of adhesion by,
\begin{equation}
    G_r = W_{\textnormal{ad}} ^r (\theta= 0 , v) = W_{\textnormal{ad}}(v)
    \label{eq7}
\end{equation}
As the crack velocity approaches to $0$, the threshold value of work of adhesion of a flat interface is $W_{\textnormal{ad}}^0$. For any finite crack velocity, $W_{\textnormal{ad}}(v) > W_{\textnormal{ad}}^0$. Thus, $G_r$ must be large in order to satisfy the condition given in equation (\ref{eq7}). As the crack length increases, ${G_r}$ reduces, resulting in a decrease in the crack velocity. 
Thus, ${W_{\textnormal{ad}}(v)}$ approaches a constant value of ${W_{\textnormal{ad}}^0}$, when the crack stops propagating, and the equilibrium crack length will be ${a = a^0}$. 
As from equation (\ref{eq4}) and (\ref{eq7}),
\begin{equation}
    W_{\textnormal{ad}} = \frac{K_I^2}{E/(1-\nu^2)}
    \label{eq8}
\end{equation}
For constant wedging displacement $h$, the energy release rate in mode $I$ is given by ~\cite{lawn1993fracture},
\begin{equation}
    G_r = \frac{3 E^{*} h^2 d^3}{4 a^4}
    \label{eq9}
\end{equation}
where $2h$ is the wedge width (thickness of the glass coverslip for our case), $d$ is sample thickness. In Fig. \ref{fig:6}(a) we can observe that initially crack front is rapidly propagating i.e. energy release rate is higher for smaller crack length and decreases with larger crack length. The behavior is in agreement with the equation (\ref{eq9}). Thus from equation (\ref{eq7}) and (\ref{eq9}), we can express the relation between work of adhesion with crack length as,
\begin{equation}
    W_{\textnormal{ad}} = \frac{3 E^{*} h^2 d^3}{4 a^4} =  \frac{K_I^2}{E^{*}}
    \label{eq10}
\end{equation}
and for equilibrium crack length $a=a_0$,
\begin{align}
    W_{\textnormal{ad}}^0 = \frac{3 E^{*} h^2 d^3}{4 a_0^4}
\end{align}
Thus, from the above expression, we can relate the interfacial work of adhesion of the sample with the equilibrium crack length.

Experimentally, the propagation of crack front and the equilibrium crack length as a function of nanoparticle concentration were investigated from the experimental setup shown in Fig. \ref{fig:4}. The underlying inverted microscope captures the crack front and records its propagation using a high speed camera. 
\begin{figure}[ht!]
\centering
\includegraphics[width=0.5\textwidth]{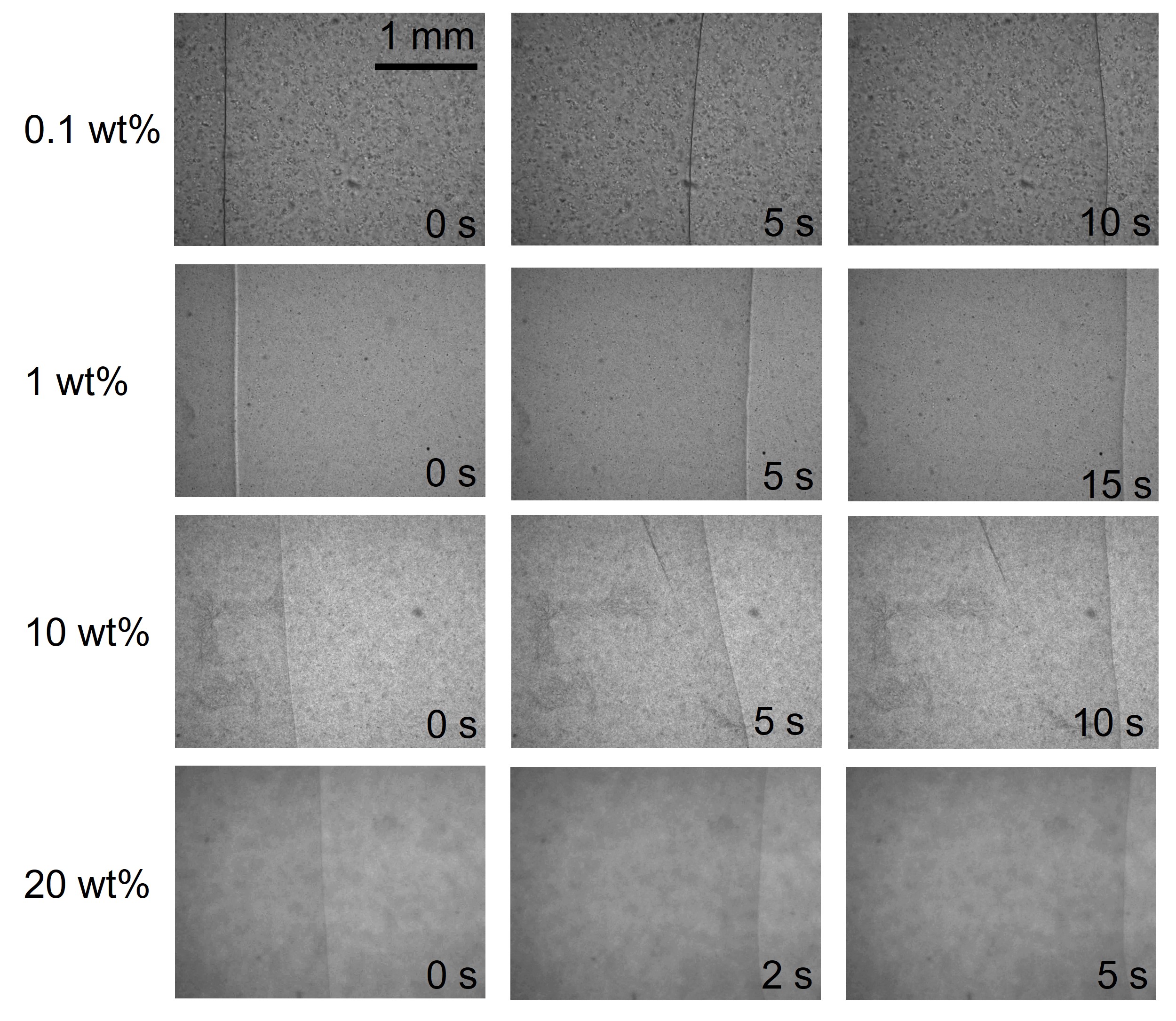}
\caption {Optical microscopic images of advancing crack front for two flat surfaces for different concentrations of nanoparticles in PDMS. Different time frames show the crack advancing through the interface with increasing time.}
\label{fig:5}
\end{figure}
Figure \ref{fig:5} shows the optical images of the crack front propagation for the nanoparticle PDMS nanocomposite samples captured from the inverted microscope. The surfaces that were studied were pure PDMS- pure PDMS, pure PDMS-PDMS(NP) (PDMS with nanoparticles), and both PDMS(NP)-PDMS(NP). Figure \ref{fig:6} shows that the dynamic and equilibrium crack length data extracted from Fig. \ref{fig:5} for various samples and combination of samples. Figure \ref{fig:6}(a) shows that with increasing nanoparticle concentration, the dynamic crack length changes in non-monotonic manner. We can also see from Fig. \ref{fig:6} (b), that the equilibrium crack length for both PDMS-PDMD(NP) and PDMD(NP)-PDMD(NP) first increases rapidly for a very small concentration of nanoparticles and then decreases exponentially with the increasing nanoparticles concentration. So the interfacial work of adhesion $W_{\textnormal{ad}}$ decreases rapidly for a very small concentration of nanoparticles (0-1\% w/w) and thereafter increases exponentially with increasing nanoparticles concentration (1-20\% w/w). The results can be understood from the theory discussed in the next section. 

In the present case, the elastic modulus of pure PDMS, is $E = 1.8 $ $\mathrm{MPa}$ and the Poisson's ratio is $\nu = 0.5$. But when the PDMS is mixed with nanoparticles, both of it's elastic modulus and Poisson's ratio changes non linearly, which is the main reason for the non-monotonic, first decrease and then increase, change in work of adhesion for both cases i.e. for PDMS-PDMS(NP) and PDMS(NP)-PDMS(NP). Hence, externally deposited nanoscale heterogeneties in pure polymers affect their mechanical and adhesion behavior as presented in this study. 

\section{Conclusion}
In conclusion, we prepared thin sheets of polymer nanocomposite, polydimethylsiloxane (PDMS) mixed with silica nanoparticles to study the role of externally deposited surface heterogeneties on their adhesion behavior. The prepared sheets had homogeneous nanoparticle, as analyzed by the bright field (BF) and dark field (DF) images. With increasing nanoparticle concentration, the transparency of the sheets decreased, which further confirm the uniform dispersion of nanoparticles in the polymer matrix. Adhesion between two surfaces was measured via wedge test, where a thin glass slide was inserted between two sheets and the dynamic and equilibrium crack length was measured. We observed very non-linear and non-monotonic change in the equilibrium crack length as a function of nanoparticle concentration. This further confirms that the mechanical, and hence the adhesion properties of polymer nanocomposite sheets vary in highly non-linear manner. 

\bibliography{Manuscript}

\end{document}